\documentclass{article}

     \usepackage[nonatbib]{neurips_2020}

\usepackage[utf8]{inputenc} 
\usepackage[T1]{fontenc}    
\usepackage{hyperref}       
\usepackage{url}            
\usepackage{booktabs}       
\usepackage{amsfonts}       
\usepackage{nicefrac}       
\usepackage{microtype}      

\usepackage{mathtools}
\newtheorem{theorem}{Theorem}[section]
\newtheorem{definition}[theorem]{Definition}
\newtheorem{example}[theorem]{Example}

\newcommand{\chapterref}[1]{\hyperref[ch:#1]{Chapter~\ref{ch:#1}}}

\newcommand{\claimref}[1]{\hyperref[claim:#1]{Claim~\ref{claim:#1}}}

\newcommand{\corollaryref}[1]{\hyperref[cor:#1]{Corollary~\ref{cor:#1}}}

\newcommand{\definitionref}[1]{\hyperref[def:#1]{Definition~\ref{def:#1}}}

\newcommand{\equationref}[1]{\hyperref[eq:#1]{Equation~\ref{eq:#1}}}

\newcommand{\factref}[1]{\hyperref[fact:#1]{Fact~\ref{fact:#1}}}

\newcommand{\figureref}[1]{\hyperref[fig:#1]{Figure~\ref{fig:#1}}}

\newcommand{\tableref}[1]{\hyperref[tab:#1]{Table~\ref{tab:#1}}}

\newcommand{\itemref}[1]{\hyperref[item:#1]{Item~(\ref{item:#1})}}

\newcommand{\lemmaref}[1]{\hyperref[lem:#1]{Lemma~\ref{lem:#1}}}

\newcommand{\propref}[1]{\hyperref[prop:#1]{Proposition~\ref{prop:#1}}}

\newcommand{\propositionref}[1]{\hyperref[prop:#1]{Proposition~\ref{prop:#1}}}

\newcommand{\remarkref}[1]{\hyperref[rem:#1]{Remark~\ref{rem:#1}}}

\newcommand{\sectionref}[1]{\hyperref[sec:#1]{Section~\ref{sec:#1}}}

\newcommand{\theoremref}[1]{\hyperref[thm:#1]{Theorem~\ref{thm:#1}}}

\newcommand{\Psymb}{\mathbb{P}}

\DeclareMathOperator*{\ProbOp}{\Psymb r}
\renewcommand{\Pr}{\ProbOp}

\newcommand{\Prob}[1]{\Pr\left[ #1 \right]}

\newcommand{\A}{{\cal A}}

\newcommand{\X}{{\cal X}}

\newcommand{\defeq}{\stackrel{\small towards\mathrm{def}}{=}}

\newcommand{\R}{\mathbb{R}}
\newcommand{\I}{\mathbb{I}}

\usepackage{bm}

\newcommand{\ignore}[1]{}

\renewcommand{\epsilon}{\varepsilon}

\title{Towards General-purpose Infrastructure for \\Protecting Scientific Data Under Study}

\author{
  Andrew L. Trask \\
  OpenMined\\
  University of Oxford\\
  \texttt{liamtrask@gmail.com} \\
  \and
  Kritika Prakash\\
  OpenMined\\
  IIIT Hyderabad\\
  \texttt{kritika.prakash@research.iiit.ac.in} \\
}

\begin{document}

\maketitle

\begin{abstract}
  The scientific method presents a key challenge to privacy because it requires many samples to support a claim. When samples are commercially valuable or privacy-sensitive enough, their owners have strong reasons to avoid releasing them for scientific study. Privacy techniques seek to mitigate this tension by enforcing limits on one's ability to use studied samples for secondary purposes. Recent work has begun combining these techniques into end-to-end systems for protecting data. In this work, we assemble the first such combination which is sufficient for a privacy-layman to use familiar tools to experiment over private data while the infrastructure automatically prohibits privacy leakage. We support this theoretical system with a prototype within the Syft privacy platform using the PyTorch framework.
\end{abstract}

The scientific method has become one of humanity's most successful and fundamental sources of truth and our chief weapon of innovation and prosperity \cite{rossi2013francis, psillos2005scientific}. However, it has a core constraint, it relies on the ability to observe the subject we want to understand \cite{ sep-scientific-method}. Despite the so-called ``big data revolution'', not all data is universally available \cite{data_sharing_doesnt_happen}. Specifically, the more personal and/or valuable the hypothesis we wish to validate, the more personal and/or valuable the data it requires (i.e., if an output is sensitive or valuable so too is the input capable of creating it). This creates a dilemma for holders of particularly personal and valuable data; Not only does every customer instantly becomes a competitor for all uses of the data (even and especially within academia), but every act of sharing carries with it significant ethical and legal risk. \cite{ascoli2015sharing, data_sharing_doesnt_happen, ascoli2017win, gal2019competitive, ethics_of_big_data}. These combined effects polarize the data marketplace such that data is either shared as rapidly as possible or not at all. 

While the former consequence is perhaps more well known, the latter is particularly unfortunate. The inability to analyze society's most valuable and personal data is the inability to answer life's most valuable and personal questions - perhaps the questions most worth answering. In this work, we argue that the key to solving this market failure is the creation of an end-to-end system for the automatic protection of data under study - such that data can be studied without being shared and without data owners actively participating in experiments. We propose such a system and design an open source prototype.

\paragraph{Technical Guarantees for Protecting Data Under Analysis} The central puzzle of privacy is that we need to share information to collaborate but we cannot limit our collaborators from using such information against us. Privacy enhancing technologies seek to allow information to be collaboratively used in a way that its future use can be carefully limited by the data owner. In the context of empirical research, we observe that these technologies provide three guarantees involving two personas: the data scientist and the data owner. These three guarantees are: protecting data from being copied by the data scientist, preventing statistical queries/results from being copied by the data owner, and preventing such statistical techniques from memorizing data in a way that the original data could be later extracted from it. With some exceptions, privacy enhancing technologies can be neatly grouped into these three categories when used in the context of empirical research.

Preventing a data scientist from copying and subsequently using data can be accomplished by bringing the algorithm to the data via federated learning/analytics \cite{fl_paper, fl_at_scale}. Preventing a data owner from copying and subsequently using the algorithm sent to the data can be accomplished using any flavor of encrypted computation: homomorphic encryption, secure multi-party computation, and functional encryption \cite{bogdanov2014input, barbosa2012delegatable}. Preventing the statistical computation from memorizing data can be accomplished through output privacy, most notably differential privacy \cite{dwork2006calibrating, dwork2006our, mironov2017renyi, dwork2016concentrated, abadi2016deep, papernot2016semi}. If combined properly, a system with such a combination of integrated techniques could provide end-to-end guarantees sufficient for general data science over private data.

\subsection{System Requirements}
\label{sect:requirements}
Despite tremendous progress on the theoretical capabilities of these systems, no implementation has yet emerged as a general-purpose alternative sufficient for scientists at large to no longer aggregate data (see supplementary material for a comprehensive overview of existing projects). We assert that such a system would require the following integrated features:

\begin{itemize}
  \item \textbf{RPC (Federated Learning):} allow one to work with data on machines they do not control.
  \item \textbf{Arbitrary Pre-publish and Post-publish Differentail Privacy Composition:} facilitate efficient tracking of arbitrary remote computation both before (entity l2-norm) and after (composition) variables are made public \cite{near2019duet, wang2019subsampled}.
  \item \textbf{User-level Permissions:} require certain privacy budget constraints to be met before statistical results can be released to a data scientist user.
  \item \textbf{Adaptive Budgeting, Filter, and Approximate Odometer:} allow arbitrary exploration of the data while informing/limiting the data scientist based on how much budget remains \cite{feldman2020individual}.
  \item \textbf{Budget Simulations:} allow for remote analysis to occur which tracks a hypothetical budget so that a data scientist can measure whether, at the end of many compositions, the accuracy gained from their model would be worth the privacy budget spend and, if it does not, decide not to download any of the results (actually spend the budget).
  \item \textbf{Individual DP:} track privacy at the individual level for various reasons. Chief among them is the need for individuals represented in multiple datasets (perhaps even at multiple institutions) to ensure that there is an upper bound on the total amount of unique statistical information released about them. This is what actually prevents harm \cite{ebadi2015differential}.
\end{itemize}
    
Nearly all of these properties exist in at least one tool or theoretical contribution. The exception is sufficiently automatic sensitivity. Our first major contribution is a general pre-publish composition language (flexible enough for any polynomial function) followed by a proposal for how it can be combined with previously proposed components to create an end-to-end system with these attributes. We finish with empirical baselines measuring the performance of a prototype system.

\section{PrivateScalar}
\label{sec:privatescalar}

We propose a novel tool for sensitivity analysis which models a database query as a polynomial over scalar values. Each free variable corresponds to an input variable contributed from a unique entity. 

\begin{definition} [PrivateScalar]
\label{def:private-scalar}
Let $y$ be a private scalar constructed using inputs from $n$ entities formed with the following metadata (Private variables are $\textbf{bold}$):

\begin{itemize}
    \item[$y^g:$]($\R^{n} \rightarrow \R$) a polynomial function with a single, clipped indeterminate for each entity contributing to $\textbf{y}$. 
    \item[$\textbf{y}^x$] : the vector $\textbf{y}^x \in \R^n$ represents the underlying value of each indeterminate within $y^g$ which, if input to the polynomial, returns $\textbf{y}^g(\textbf{y}^x) = \textbf{y}$. 
    \item[$y^f$] : (floor) each element of the vector $y^f \in \R^n$ represents the minimum possible value of the indeterminate $y^x_i$ input into $y^g$.
    \item[$y^c$] : (ceiling) each element of the vector $y^c \in \R^n$ represents the maximum possible value of the indeterminate $y^x_i$ input into $y^g$. 
    \item[$\textbf{y}$] : the value of the private scalar, taken by clipping each indeterminate $\textbf{y}^x$ within the range of $y^c$ and $y^f$ and passing it into the polynomial $y^g$
\end{itemize}
\end{definition}

Let us consider an example. Consider 100 individuals each contributing their age to a study. Entity $i$ would initialize a PrivateScalar with an internal polynomial with 100 indeterminates, all with factors equal to 0 except for the $i^{th}$ factor which is 1. Similarly, $y^x$ would also be a one-hot vector, with entity $i$'s age represented in the $i^{th}$ position. Executing $\textbf{y}^g(\textbf{y}^x)$ would thus simply return entity $i$'s age. However, if one wished to compute any arbitrary function over the 100 ages (such as a sum, mean, or arbitrary polynomial), each operation would manipulate the polynomial over inputs instead of manipulating the inputs themselves. This keeps the inputs disentangled as operations construct a complex query, the result of which is only calculated when the result is to be published (with noise).

We now consider how PrivateScalar can be applied within the context of the individual R\'enyi DP of \cite{feldman2020individual}. Example \ref{def:lipschitz-example} from \cite{feldman2020individual} shows how to determine the entity-specific epsilon spend per query in the context of a Lipschitz function over a database (See appendix or \cite{feldman2020individual} for the definition of individual RDP using notation matching this example).

\begin{example}[Lipschitz analyses]
\label{def:lipschitz-example}
Suppose that $g:(\R^d)^n\rightarrow \R^{d'}$ is $L_i$-Lipschitz in coordinate $i$. For  $\phi\colon \X \to \R^d$, let $\A(S) = g(\phi(X_1),\dots,\phi(X_n)) + \xi$, $\xi\sim N(0,\sigma^2 \I_{d'})$. Assume that for some $X^\star$, $\phi(X^\star)$ is the origin.
By using $X^\star$ to replace a removed element (namely $S^{-i} = (X_1,\dots,X_{i-1},X^\star,X_{i+1},\dots,X_n)$), then  $\A$ satisfies $\left(\alpha,\frac{\alpha L_i^2 \|\phi(X_i)\|_2^2}{2\sigma^2}\right)$ individual RDP for $X_i$.
\end{example}

Given that $g$ is a query over private data $X$ (where each element of $X_i$ comes from unique entity $i$), the key question this example answers is: how much privacy budget does entity $i$ spend when the output of $\A(S) = g(\phi(X_1),\dots,\phi(X_n)) + \xi$ is made public? The answer to this question is $\left(\alpha,\frac{\alpha L_i^2 \|\phi(X_i)\|_2^2}{2\sigma^2}\right)$, the two key terms of which are $L_i^2$ and $\|\phi(X_i)\|_2^2$, which PrivateScalar reveals. PrivateScalar mirrors this definition. $g$ corresponds to $y^g$, an arbitrary polynomial over $y^x$, which corresponds directly to $X$. $\phi()$ is expressed within the factors of the polynomial. Thus, recovering the two key terms $L_i^2$ and $\|\phi(X_i)\|_2^2$ are as simple as considering the Lipschitz bound on each free variable in the polynomial and the L2 norm of the input respectively.

\paragraph{Recovering The Lipschitz Constant} Any part of the query which involves information mixing (non additively) between multiple entities is captured in $g$.  The term $L_i^2$ refers to the squared Lipschitz constant of $g$, with respect to the output of $\phi(X_i)$. This Lipschitz constant is not over an infinite range, however, but is only over the range of values expressed by the removal/replacement of $\phi(X_i)$. 

However, while the L2 norm of the input is easy to compute, the Lipschitz bound on $g$ can be more complex, because the derivative of $y^g$ with respect to $X_i$ may be conditioned on private data from entities other than $i$ (if $y^g$ multiplies two or more free variables). In the literature it is often assumed that each entity is able to see the remaining budget with respect to itself, thus each individual's budget should only be conditioned on private  data of itself (and public data from others)\footnote{It is unclear whether this is a problem in all settings - as each entity need not necessarily know the current budget with respect to their data, and epsilon is considered private in \cite{feldman2020individual}. We offer a conservative alternative for sake of generality.} 

To avoid this, challenge we instead consider the maximum possible derivative over all possible $y^g$ polynomials given the publicly known ranges of all of its free variables (as defined by $y^f$ and $y^c$, which are public). While computing this derivative can be challenging for complex polynomials, some special cases exist.

\paragraph{Special Cases} In the case that the polynomial is comprised exclusively of non-negative factors and free variables, the polynomial becomes positively monotonic. In this case, finding the maximum possible derivative can be computed in closed form by considering $y^g$ when all inputs $y^x$ equal exactly their upper bounds $y^c$. In the case that $y^g$ is a first-degree polynomial, then the Lipschitz constant is equal to the public coefficient corresponding to $y^x$.

Thus, for any private scalar we seek to publish, which may have been formed from any polynomial over entity information, the amount of privacy spend for each contributing entity can be determined. This makes PrivateScalar comparable to previous work in sensitivity type systems, with several advantages over prior works (see Table \ref{tab:dp-tools} for prior works):
\begin{itemize}
\item in the spirit of \cite{feldman2020individual}, our sensitivity system employs some data-dependent information (namely $x_i$), which is tighter than previous purely data-independent approaches.
\item this tool is capable of private-private multiplication between, and non-linear functions over, private values - which (to our knowledge) no existing tool for sensitivity analysis can bound.
\end{itemize}

The primary constraint of the PrivateScalar data-structure is one of performance; if one performs many computation on PrivateScalar values which involve multiple entities and potentially negative values, the underlying polynomial is likely to become very large, and the Lipschitz bounds expensive to compute.

\section{An End-to-End System for Private Data Analysis}

To fulfill the requirements of section \ref{sect:requirements}, we combine PrivateScalar with the following tools toward an end-to-end system.

\begin{itemize}
  \item \textbf{RPC (Federated Learning):} we leverage the RPC federated learning capabilities of the PySyft PPML framework.
  \item \textbf{Arbitrary Pre-publish and Post-publish Composition:} pre-publish composition is accomplished via PrivateScalar, and post-publish composition via the autodp tool of \cite{wang2019subsampled}.
  \item \textbf{Permissions:} We integrate with the new 0.3.0 alpha release of PySyft, whose RPC framework has object-level, user-level permissions.
  \item \textbf{Adaptive DP Filter and Approximate Odometer:} we adopt the approach of \cite{feldman2020individual}.
  \item \textbf{Budget Simulations:} a data scientist can also copy their current odometer (remaining privacy budget) into a simulated data structure, then using this copy to simulate how a budget could be sent by telling this simulated odometer that it is publishing objects that haven't yet actually been downloaded. In this way, a data scientist can plan how a budget could be spent and search for the optimal privacy/accuracy tradeoff for their system before actually spending the true budget (downloading teh results).
  \item \textbf{Individual DP:} we augment the autodp framework of \cite{wang2019subsampled} with the individual DP method proposed in \cite{feldman2020individual} within the PySyft framework.
\end{itemize}

While length will not allow a full exposition of each of these features, perhaps the most important integration to discuss is that between the permissions system of PySyft and the privacy budgeting mechanisms proposed above. Importantly, this means that a data owner can set a privacy budget for a data scientist, and the data scientist can perform any analysis they desire (and download results) as long as they stay under their budget. The combination between PrivateScalar and autodp within an existing statistical tool (PySyft + PyTorch) is, we argue, an important threshold in facilitating arbitrary data science over private data such that the data owner may rely heavily on automation to protect their information.

\section{Conclusion and Future Work} 

In this work we survey recent techniques to propose a system sufficient to achieve an important milestone for the broader scientific community: the ability for a non-technical data owner to allow a data scientist to safely perform arbitrary data analysis over their private information such that the infrastructure can automatically protect the data under study (where ``protect'' is defined by a privacy budget). While most components for our proposed system exist in recent work, we identify and propose an important missing piece, online, arbitrary sensitivity (pre-publish) composition, and propose a tool capable of satisfying it. Finally, we argue that when combined with previous work in the way we recommend, the protection of private data while under (somewhat) arbitrary data analysis crosses an important usability threshold: automation. We present our prototype implementation as an open-source tool for further maturation\footnote{Prototype will be shared after blind review is complete.}. Future work will focus on shoring up side-channel attacks, increasing computational performance, tightening differential privacy bounds, extending PrivateScalar beyond polynomials to threshold functions, and integrating more closely with other encrypted computation techniques offered by PySyft.

\bibliography{iclr_conference}
\bibliographystyle{plain}
\pagebreak
\section*{Appendix}

\section{Protect Data From Being Copied}

The scientific method requires a scientist to collect empirical evidence to support their claim. However, every popular statistical software tool assumes that its user has copied the information they seek to study onto computational resources they control. The natural consequence of this assumption is that scientists are technically capable of using such data for any purpose they desire - including uses the original data owners my oppose. Their data is, both literally and figuratively, ``out of their hands''.

\paragraph{Theory} The solution is simple but cumbersome. Instead of aggregating data to a single location, a scientist should delegate their analysis to data owners to run on their own hardware, thus avoiding the need to obtain a copy of the data they study. A huge burst of research has ensued in this direction seeking to break algorithms into discrete parts which can be run by multiple parties in this way. Federated learning (FL) refers such work in a machine learning setting and federated analytics to statistics more broadly \cite{fl_paper, fl_at_scale}.

\paragraph{Implementations}While much progress has been made in studying how to efficiently break algorithms apart, and many systems for FL are being built, no implementation has yet emerged as a general-purpose alternative sufficient for scientists at large to no longer aggregate data. 

\begin{table}[]
\begin{tabular}{ccccccccccccc}
\hline
\multicolumn{1}{|c}{Name} & MoYr  & Make    & Base  & RPC & OPC & AM & DP  & PM & M & B & S & \multicolumn{1}{c|}{T} \\ \hline
PySyft                    & Jul17 & OpenMnd   & TH    & Y   & Y   & Y  & 3rd & Y  & Y & Y & Y & Y                      \\
TFF                       & Sep18 & Google  & Any   & N   & N   & N  & 3rd & N  & N & N & Y & N                      \\
FATE                      & Sep18 & WeBank  & TH,+  & N   & N   & N  & 3rd & N  & N & N & Y & ?                      \\
LEAF                      & Dec18 & CMU     & TF    & N   & N   & N  & 3rd & N  & N & N & Y & ?                      \\
eggroll                   & Jul19 & WeBank  & TF, + & ?   & N   & ?  & 3rd & N  & N & N & Y & ?                      \\
PaddleFL                  & Sep19 & Baidu   & PD    & Y   & N   & ?  & Y   & N  & N & N & Y & ?                      \\
FLSim                     & Nov19 & iQua    & TH    & N   & N   & N  & N   & N  & N & N & Y & N                      \\
Clara                     & Dec19 & NVIDIA  & TF    & N   & N   & Y  & Y   & N  & N & N & Y & ?                      \\
IBMFL                     & Jun20 & IBM     & KS+   & N   & N   & N  & Y   & N  & N & N & Y & Y                      \\
FLeet                     & Jun20 & EPFL    & TF?   & N   & N   & ?  & Y   & N  & A & N & ? & ?                      \\
IFed                      & Jun20 & WuhanU   & CS    & N   & N   & N  & Y   & N  & N & N & Y & Y                      \\
FedML                     & Jul20 & FedML   & TH    & Y   & N   & N  & Y   & N  & A & N & Y & Y                      \\
Flower                    & Jul20 & Cmbrdge & Any   & Y   & N   & N  & 3rd & N  & Y & ? & Y & Y                      \\ \hline
\end{tabular}
\caption{Federated Learning systems listed in order of publication (earlier of paper or Github repo). Name: the name of the system. MoYr: the month and year of publication. Make: the sponsoring organization. Base: the primary ML framework (TH=PyTorch, TF=Tensorflow, PD=Paddle, KS=Keras, CS=Custom, Any=Arbitrary, += multiple truncated for space). RPC: can a data scientist / coordinator node push jobs to workers or do workers pull them? OPC: Object level RPC - can a data scientist / coordinator interactively control arbitrary objects on the data nodes. AM: Can a data scientist / coordinator set/change the model architecture being trained without having access to client workers (or having to restart them)? DP: Does the framework natively support some kind of Differential Privacy (3rd = third party library can support). PM: Does the framework have a permissions system such that the data scientist/coordinator is considered a malicious adversary where the infrastructure's job is to prevent them from using their access to steal private data. M: Does the framework have mobile support (A=Android only)? B: Can the framework run in the browser? S: Can the framework run on servers? T: Can the framework run on IoT devices? \cite{Ryffel2018AGF, TFF, FATE, LEAF, eggroll, paddlefl, clara, ludwig2020ibm, damaskinos2020fleet, cao2020ifed, he2020fedml, beutel2020flower, wang2020optimizing}}
\label{tab:fl-libs}
\end{table}

To the knowledge of these authors, all systems for FL require a scientist to actively coordinate with each data owner on every experiment (or with the builders of their software)\cite{kairouz2019advances}. If they seek to ensure their data is safe, the data owner must, for each experiment, read the code of each experiment for themselves (which assumes the data owner has the time and expertise to do so)\cite{kairouz2019advances}. Current FL infrastructure is high-touch, high-trust, high-expertise, and primarily suited for enterprises or mobile applications looking to do the same experiment over a long period of time with parties who trust their intentions (or don't really have a choice)\cite{kairouz2019advances}.

\section{Protect Statistical Models From Being Copied}

Setting aside open problems in federated learning, even idyllic FL still requires each data scientist to divulge their statistical techniques (and their work-in-progress model) to the data owner for training\cite{kairouz2019advances}. While the chief concern of privacy technology is the protection of personal information, in order for the protections of federated learning to be viable in a competitive marketplace, scientists (whether academic or commercial) need some ability to prevent data owners from taking advantage of their access to the statistics being created. The field of encrypted computation promises to address this.

\paragraph{Theory} Encrypted computation, often called ``input privacy'', allows for multiple parties to jointly compute a function without revealing their respective inputs to each other \cite{bogdanov2014input}. Within the field of cryptography, this field is called Secure Multi-party Computation (SMPC), special cases of which include Homomorphic Encryption and Functional Encryption \cite{barbosa2012delegatable}. Secure hardware can also provide encrypted computation as an alterantive to SMPC \cite{costan2016intel}. While all of these options can offer Turing complete encrypted computation, the ideal technique (or combination) for any particular application varies based on the available compute, ram, and network infrastructure.

\paragraph{Implementations} Recent work has produced a flurry of libraries and chips for general purpose encrypted computation, particularly integer-level encrypted computation which is the preferred variety for statistical computation\cite{costan2016intel, chen2017simple}. Several tools exist which augment existing data science tools to run in an encrypted state \cite{gunning2019crypten, dahl2018private}. Furthermore, the long-standing challenge of providing encrypted computation frameworks which are performant on non-trivial machine learning tasks has, in general, been accomplished, although encrypted CPU training is typically still 10x+ slower than plain-text CPU training \cite{wagh2020new}. Creating more performance algorithms and implementations is still a very active area of research \cite{wagh2020new}. 

However, existing implementations fall short in a way similar to federated learning. Namely, no framework yet exists which is capable of tracking the encrypted computation while it is happening such that a data owner can automatically prevent data from simply being copied and sent to the data scientist during the computation process. And because the computation is encrypted, it's even more challenging for a data owner to read and understand the code they are running on their sensitive data (they would need to be an encrypted computation expert, of which there are very few). 

For both federated learning and encrypted computation implementations, the missing piece is a security policy (linked to a user permissions system) which can actively and automatically ensure that each user of a system doesn't ever learn too much about the underlying data.

\subsection{Prevent Statistical Models from Memorizing Data}

The dominant solution for preventing statistical models from memorizing data is \textit{differential privacy}. Introduced as a privacy constraint around database queries (simple statistics), it has been generalized to offer similar protections over even the most complex statistical analysis - ensuring that statistical results don't compromise the privacy of the records they describe.

\paragraph{Theory}
Proposed by \cite{dwork2006calibrating, dwork2006our}, differential privacy builds on the intuition that a query from a database is privacy preserving if removing or replacing any entry in the database doesn't change the result of the query. When a query's result does change given input perturbations, various ``randomized algorithms'' have been proposed to add noise to a database query in such a way that rigorous, worst-case bounds can be set on the probability that an input data-point could be inferred from the query result. Several popular modifications of the original DP definition have been proposed - for which many special-purpose mechanisms have been designed to find the best privacy/accuracy trade-off for specific algorithm types\cite{mironov2017renyi, dwork2016concentrated, abadi2016deep, papernot2016semi}.
 
\paragraph{Epsilon-delta DP} Following the notation found in \cite{feldman2020individual} , let $S=(X_1,\dots,X_n)$ be an analyzed dataset, and $S^{-i} \defeq (X_1,\dots,X_{i-1},X_{i+1},\dots,X_n)$ be the analyzed dataset after removing point $X_i$. $(\epsilon,\delta)$-differential privacy (DP) measures privacy leakage using $\epsilon$, an upper bound on the distance between queries on $S$ and $S^{-i}$, and $\delta$, the probability that the measure of distance fails.

\begin{definition}
\label{def:dp}
A randomized algorithm $\A$ is $(\epsilon,\delta)$-DP if for all datasets $S = (X_1,\dots,X_{n})$,
$$\Prob{\A(S) \in E}\leq e^\epsilon\Prob{\A(S^{-i}) \in E} + \delta,\text{ and } \Prob{\A(S^{-i})\in E}\leq e^\epsilon\Prob{\A(S)\in E} + \delta,$$
for all $i\in[n]$ and all measurable sets $E$ \cite{dwork2006calibrating,dwork2006our,feldman2020individual}.
\end{definition}

\paragraph{Privacy Budgeting} Each statistical query spends a certain degree of $\epsilon$. In theory, the total amount of $\epsilon$ a data scientist is allowed to acquire in their analysis is called a \textit{privacy budget}. If available, a data scientist can track their current spend of the budget using a \textit{privacy odometer} for the exact value or a \textit{privacy filter} to simply indicate whether or not the budget (or some partial threshold of it) has been exceeded.

\paragraph{Adaptivity and Scope} However, the ability to measure a privacy budget by composing multiple rounds of privacy parameters $\epsilon$ and $\delta$ is very complex. The simplest and earliest composition theorems are non-adaptive and global, meaning that a data scientist must know all of the queries they want to run before viewing the output of the first one (no interactive exploring of the data), and the privacy budget refers to the leakage from the entire dataset as a whole (as opposed to individuals within the dataset whom we seek to protect). While the former might seem obviously problematic, the latter requires context.

Consider two data scientists, Bob and Alice, running their experiments against two medical data-sets at two different hospitals (respectively). Importantly, these data-sets have 
overlapping patients even though they're at different hospitals. If Bob and Alice are each given privacy budgets of $\epsilon = 3$, if they compare their results they could (in theory) learn more than $\epsilon = 3$ worth of information about the overlapping patients in their dataset! 

That is to say, just limiting each scientist to a certain amount of $\epsilon$ budget doesn't ensure that people in the dataset are actually protected if scientists share results (or simply make them public). Instead, we should limit each \textit{data subject} (in this case medical patients) to a certain amount of budget, no matter where that budget is being spent. DP which tracks a unique epsilon per individual is called individual DP.

\paragraph{Adaptive, Individual, Renyi-DP} While a full survey of adaptive and individual DP methods is out of the scope of this work, we do focus on a particular extension of $(\epsilon,\delta)$-DP called R\'enyi DP-(RDP), which was recently extended with useful definitions and examples for adaptive composition with individual differential privacy by \cite{feldman2020individual} and \cite{ebadi2015differential}. It is upon this work we will propose our end-to-end system.

\subsection{Implementations}

Recent work has seen many new tools for differential privacy, motivated by the desire for DP's complex analysis to be conveniently available both to practitioners and laymen alike. Works can be split into two groups by deployment environment: database differential privacy (\cite{mcsherry2009privacy}, \cite{proserpio2012calibrating}, \cite{johnson2018towards}, \cite{dpella}, \cite{whitenoise}) and differential privacy within more general software programs: MapReduce (\cite{roy2010airavat}), functional programs (\cite{gaboardi2013linear, haeberlen2011differential, near2019duet}), federated datasets (\cite{narayan2012djoin}), and Python programs (\cite{mohan2012gupt, wang2019subsampled}). 

Of the latter tools which support DP over arbitrary functions, there are two subgroups. Most tools exclusively focus on tracking and composing $\epsilon$ (post-query composition analysis), but some tools also track the arbitrary computation occurring before a publish event so that noise can be automatically calibrated to meet a certain budget \cite{near2019duet, reed2010distance, gaboardi2013linear, azevedo2018, barthe2015higher, zhang2019fuzzi, dpella}. Specifically, such tools track the ``sensitivity'' of a function's output to input perturbations. It is noteworthy, however, that some DP techniques calibrate noise based on measures other than sensitivity, but no automatic tools yet leverage them\cite{feldman2020individual}.

\paragraph{Relationship to Permissions Systems} Similar to systems for federated learning and encryption, in our view the primary shortcoming in most systems for tracking DP is that nearly all systems capable of general purpose data science are decoupled from a remote-procedure call and object permissions system.

Put another way, while tools for analysis can tell you if your statistical analysis would leak private information if published, they can rarely use the conclusion of such analysis to explicitly prevent you from publishing anyway. The primary exception can be found within some database-query style DP tools, but these lack the ability to do arbitrary computation. A standout exception is the early work of \cite{mohan2012gupt} which does allow arbitrary programs and enforces a privacy budget against a data scientist adversary.  

\paragraph{The Expressiveness of Automatic Sensitivity Tools} However, while \cite{mohan2012gupt} does enforce a privacy budget for arbitrary programs, it does so by requiring the data analyst

While some hybrid sensitivity-composition tools exist the sensitivity analysis they offer is primarily linear and entity generic. To the best of the authors knowledge, a sensitivity analysis tool does not yet exist with supports analysis over the multiplication of two private values (without the result having infinite sensitivity) or a general purpose mechanism for tracking the sensitivity of non-linear functions. Additionally, no hybrid sensitivity-composition analysis tools are in a language commonly used for data science and by extension, none have been integrated into a popular statistical tool for general-purpose use.

\begin{table}[]
\begin{tabular}{ccccccccccccc}
\hline
\multicolumn{1}{|c}{Name}    & Based                                           & RPC   & ORPC & FSA & ISA & AC   & IDP & RDP & PM  & ML  & \multicolumn{1}{c|}{UAPI}  \\ \hline
PINQ    & -              & SQL   & No   & Yes & No  & API  & No  & No  & Yes & No  & C\#   \\
Airavat & -                  & MapR  & No   & No  & No  & Map  & No  & No  & Yes & Yes & Java  \\
Reed    & -                 & Cust & No   & Yes & Yes & ? & No  & No  & No  & Yes & Cust \\
Fuzz    & PINQ     & SQL   & No   & Yes & Yes & API  & No  & No  & Yes & No  & C\#   \\
GUPT    & -                    & Pyth & No   & Yes & No  & Map  & No  & No  & Yes & Yes & Pyth \\
wPINQ   & PINQ     & SQL   & No   & Yes & No  & API  & No  & No  & Yes & No  & C\#   \\
DJoin   & -                 & SQL   & No   & Yes & No  & No   & No  & No  & Yes & No  & SQL   \\
DFuzz   & Fuzz            & SQL   & No   & Yes & Yes & API  & No  & No  & Yes & No  & C\#   \\
RAPOR  & -               & No    & No   & Yes & No & API  & No  & No & No  & No & C++ \\
HOARe   & DFuzz             & SQL   & No   & Yes & Yes & API  & No  & No  & Yes & Yes & C\#   \\
FLEX    & -               & SQL   & No   & Yes & No  & No   & No  & No  & Yes & No  & SQL   \\
Proclo  & -               & No    & No   & Yes & No & API  & No  & No & No  & No & C++ \\
Fuzzi   & aRHL               & No    & No   & Yes & Yes & API  & No  & Yes & No  & Yes & Cust \\
Duet    & -                     & No    & No   & Yes & Yes & ? & No  & Yes & No  & No  & Hskl  \\
TFpriv  &    -            & No    & No   & No  & No  & No  & No  & Yes & No  & Yes & Pyth \\
pyvacy  &    -            & No    & No   & No  & No  & No  & No  & Yes & No  & Yes & Pyth \\
autodp  &   -          & No    & No   & No  & No  & Yes  & No  & Yes & No  & Yes & Pyth \\
WNoise  &   -          & SQL    & No   & Yes  & Yes  & Yes  & No  & Yes & No  & No & Pyth \\
GoogDP  &   -          & No    & No   & No  & No  & No  & No  & No & No  & No & R,Go \\
PyDP  &   GoDP          & No    & No   & No  & No  & No  & No  & No & No  & No & Pyth \\
SwftDP  &   GoDP          & No    & No   & No  & No  & No  & No  & No & No  & No & Swft \\
dp.js  &   GoDP          & No    & No   & No  & No  & No  & No  & No & No  & No & JS \\
JavaDP  &   GoDP          & No    & No   & No  & No  & No  & No  & No & No  & No & Java \\
ClojDP  &   GoDP          & No    & No   & No  & No  & No  & No  & No & No  & No & Cloj \\
diffPR  &   GoDP          & No    & No   & No  & No  & No  & No  & No & No  & No & R \\
Opacus  &    -            & No    & No   & No  & No  & No  & No  & Yes & No  & Yes & Pyth \\
\hline
\end{tabular}
\caption{Differential privacy systems listed in order of publication year. Name:the name of the system. RPC: What language defines a query if the system allows remote operation. ORPC: Does the framework support object-level RPC (Yes) or just a static function API (No)? FSA: Can the system infer the sensitivity of supported functions (Yes) or does it need to be specified by the system designer (No)? ISA: Can FSA happen with knowledge of the underlying object sensitivity (i.e., is the sensitivity fixed per function (No) or dynamic to the object (Yes) over which the function is being called?). AC: How flexible is computation? (MAP: can pass in arbitrary map functions, but not reductions, API: arbitrary computation but only to the API the data owner explicitly develops, ?: Generally a flexible paradigm but unclear because some operations are limited or generate infinite budget spend). IDP: Supports individual differential privacy? RDP: Supports R\'enyi DP? PM: Integrated with a permissions system such that the user is considered an untrusted adversary who must stay under a privacy budget. ML: API flexible enough for general-purpose machine learning? UAPI: what language does the user use to query data? \cite{mcsherry2009privacy, roy2010airavat, reed2010distance, haeberlen2011differential, mohan2012gupt, proserpio2012calibrating, narayan2012djoin, gaboardi2013linear, barthe2015higher, johnson2018towards, zhang2019fuzzi, near2019duet, wang2019subsampled, pyvacy, abadi2016deep, prochlo, rappor}}

\label{tab:dp-tools}
\end{table}





\end{document}